\documentstyle[prl,aps,epsf]{revtex}
\draft
\begin{document}
\twocolumn[\hsize\textwidth\columnwidth\hsize\csname @twocolumnfalse\endcsname

\title{Statics and Dynamics of Vortex Liquid Crystals 
} 
\author{C. Reichhardt and C.J. Olson Reichhardt} 
\address{ 
Theoretical Division and Center for Nonlinear Studies,
Los Alamos National Laboratory, Los Alamos, New Mexico 87545}

\date{\today}
\maketitle
\begin{abstract}
Using numerical simulations we examine the
static and dynamic properties of the 
recently proposed vortex liquid crystal state. 
We confirm the existence of a smectic-A phase in the absence of 
pinning. Quenched disorder can induce a smectic state even at $T=0$.   
When an external drive is applied, a variety of 
anisotropic dynamical flow states 
with distinct voltage signatures occur, including 
elastic depinning in the hard direction and 
plastic depinning in the easy direction.
We disuses the implications
of the anisotropic transport for other systems which exhibit
depinning phenomena, such as stripes and electron
liquid crystals.  
\end{abstract}
\vspace{-0.1in}
\vspace{-0.3in}

%\vskip2pc
\vskip2pc]
\narrowtext
Recently, a new state of vortex matter termed a vortex liquid 
crystal was proposed    
to occur  
in superconductors 
with anisotropic vortex-vortex interactions \cite{Carlson}. 
In such systems, the vortex lattice first melts in the soft direction,
giving rise to an intermediate vortex smectic-A state, followed at
higher temperatures by a melting into a nematic state.   
The initial 
theoretical calculations of this transition
combined an elastic model with the Lindemann
criterion for melting;
however, the validity of this approach has been called into question 
\cite{Hu,Campbell}. The Lindemann criterion does not take
into account the proliferation of dislocations 
that is likely to occur in a smectic-A state, so a numerical investigation
would be very useful both 
to determine whether the smectic-A state can occur in 
the vortex
system as well as to
examine the 
dynamics of vortex liquid crystals in the presence of disorder. 
The physics of the vortex liquid crystal state 
should be generic to the class of problems which
can effectively be modeled as a two-dimensional (2D) system of 
particles with anisotropic repulsive interactions. 
Such a system has already been physically 
realized for magnetic colloidal particles, 
where a smectic-A state was observed along with dislocations
that have preferentially aligned Burgers vectors \cite{Keim}.   
There is also considerable interest in electron liquid crystal states, 
which may arise when anisotropic interactions in
classical electron crystals give rise to smectic and
nematic states \cite{Kivelson,Dorsey}.
Evidence for such states has 
been observed in transport measurements which show 
hard and soft directions for flow \cite{Lilly,Cooper}.   

Previous studies of
vortex smectic states 
considered vortices 
interacting with  
some form of an underlying 1D periodically modulated substrate \cite{Nelson}.  
A very similar system in which smectic states have been observed
is colloidal particles interacting 
with 1D periodic substrates \cite{Leiderer}. 
Each of these systems melts into an intermediate smectic-C state. 
When the underlying substrate is disordered rather than
periodic, application of an external drive induces an
anisotropic fluctuating force that organizes the vortices into a
moving smectic state, where the dislocations in the vortex lattice
are aligned with the direction of the applied drive
\cite{Balents,Zimanyi}. 
In the case of the proposed vortex liquid crystal state, 
the anisotropy arises when the vortex
cross section becomes elliptical 
due to an anisotropic superfluid stiffness which leads to
different effective masses in the three crystalline
directions \cite{Carlson}. 
The theoretical calculations in Ref.~\cite{Carlson} were performed 
for a system with no quenched disorder;
however, 
real superconductors often contain significant amounts of random pinning.
It would be desirable to understand 
the transport properties of vortex liquid crystals 
in order to identify signatures of the liquid crystal phase and seek
new types of dynamical phenomena in these systems. 
Understanding how quenched disorder
affects an anisotropic system of repulsively interacting 
particles is also relevant to 
dynamics in electron liquid crystal states. 

To address these issues,
we consider a 2D system of $N_v$ interacting vortices
with periodic boundary conditions in the
$x$ and $y$ directions. The  
overdamped equation of motion for a single vortex $i$ is 
\begin{equation}
\eta\frac{ d{\bf R}_{i}}{dt} = {\bf f}_{i}^{vv} + {\bf f}^{T}_{i} + {\bf f}^{p}_{i} + {\bf f}^{d}_{i} 
\end{equation}
The damping constant $\eta$ is set to unity.  
The vortex-vortex interaction force is 
${\bf f}_{i}^{vv} = \sum^{N_{v}}_{j\ne i}A_{v}K_{1}(r_{ij}/\lambda){\bf {\hat r_{ij}}}$, 
where $K_{1}$ is the modified Bessel function,
which decays exponentially for large distances, 
$\lambda$ is the London penetration depth, $A_{v}$ is the 
vortex interaction prefactor, and $r_{ij}$ is the distance between
vortices $i$ and $j$.
The Bessel function 
is 
appropriate for stiff, 3D vortex lines. We have also
considered $1/r$ interaction potentials appropriate for thin film 
superconductors as well as Yukawa interaction potentials for colloidal
particles and find the same qualitative features.
The thermal force ${\bf f}^{T}_{i}$ arises from random 
Langevin kicks with the properties $<{\bf f}^{T}_{i}> = 0$ and
$<{\bf f}_{i}^{T}(t){\bf f}_{j}^{T}(t^{\prime})> = 2\eta k_{B}T \delta(t-t^{\prime})\delta_{ij}$. 
The quenched disorder ${\bf f}_{i}^{p}$
is modeled as random pinning sites 
in the form of
attractive parabolic traps of radius $r_{p}=0.2\lambda$ and strength $f_{p}$.
The Lorentz driving force from an external applied current is ${\bf f}^{d}$. 
The system size is measured in units
of $\lambda$ and the forces in terms of $A_{v}$. 
The anisotropic interactions
are introduced  
by multiplying the 
vortex-vortex interaction force in the $x$ and $y$ 
directions by a vector $(C_{x}, C_{y})$, where the anisotropy $C=C_x/C_y$.
In this work we concentrate on the case 
$C = 1/\sqrt{10}$ which 

\begin{figure}
\center{
\epsfxsize=3.5in
\epsfbox{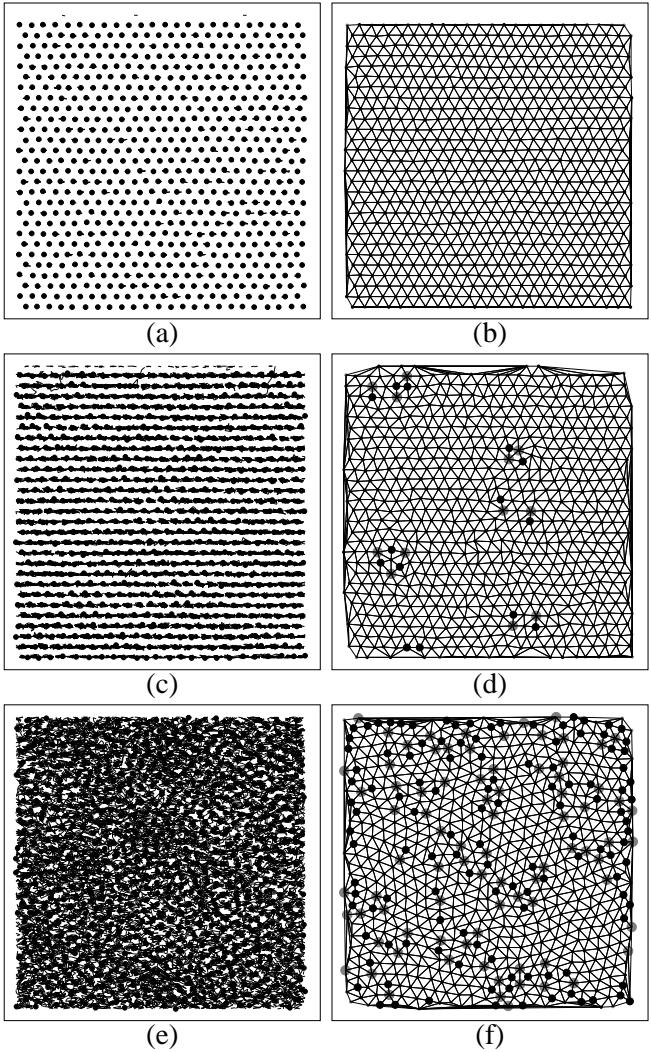}}
\caption{
(a,c,e) Black dots: vortices; black lines: vortex trajectories.   
(b,d,f) Delaunay triangulation, with topological defects (5 and 7-fold 
coordinated particles) marked as filled circles.
(a,b) $T = 0.5$;
(c,d) $T = 1.2$; 
(e,f) $T = 1.35$.
}
\end{figure}

\noindent
is the value considered in Ref.~\cite{Carlson}.  
We take the $x$ axis to be
the soft direction and the $y$ axis as the hard direction.

We first consider the case where the pinning and the external driving 
force are absent. 
In Fig.~1 we illustrate the melting of a 
$24\lambda \times 24\lambda$ system with a vortex density of 
$\rho_v=1.2/\lambda^2$.
Figure 1(a) shows
the vortex positions (dots) and trajectories (lines)
for a fixed period of time with a fixed $T = 0.5$, and
Fig.~1(b) shows a corresponding  Delaunay triangulation.
At this temperature, 
the system remains in a crystalline state with no 
dislocations. 
The vortices are undergoing 
larger random displacements in the soft ($x$) direction than 
in the hard ($y$) direction; however, there is no
long time diffusion of the particles. 
Figures 1(c) and 1(d) present the smectic-A state at $T=1.2$.
Here the trajectories have
a 1D liquid structure with 

\begin{figure}
\center{
\epsfxsize=3.5in
\epsfbox{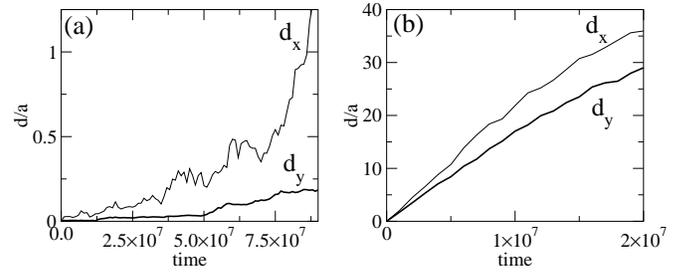}}
\caption{
The average particle displacements in each direction,
$d_{x}$ and $d_{y}$,
normalized by the lattice constant $a$, vs
time, measured in molecular dynamics steps.
(a) The smectic-A state at $T = 1.21$. 
(b) The nematic phase at $T=1.35$.
}
\end{figure}

\noindent
motion 
along the soft $x$ direction 
and no 
significant
translation of the vortices in the $y$ direction. The
Delaunay triangulation 
indicates the presence of dislocations with aligned Burgers 
vectors, which is
characteristic of the smectic-A state. 
Figures 1(e) and 1(f) illustrate the vortex liquid phase at
$T = 1.35$.
The vortex trajectories show 
clear 
diffusion in both the $x$ and $y$ directions, with more pronounced motion 
in the $x$ direction.
The dislocations are no longer aligned in a single direction, 
indicating the loss of long-range order in both
the $x$ and $y$ directions. 
These results confirm that a smectic-A state can occur 
in a system of vortices with
anisotropic interactions, as predicted by theoretical 
calculations \cite{Carlson}. We note that when the anisotropy
ratio $C$ is too small, the two-step melting transition 
illustrated here is lost.   

To further characterize the smectic state,
in Fig.~2 we plot the average particle displacements 
for the $x$ and $y$ directions, 
$d_{x} = <\sum_i^{N_v}|x_i(t) - x_i(t^{\prime})|>/N_v $ 
and  $d_{y} = <\sum_i^{N_v}|y_i(t) - y_i(t^{\prime})|>/N_v$. 
In the smectic phase at $T = 1.21$, shown in Fig.~2(a),
$d_{x}/a$ increases much more rapidly than $d_{y}/a$, 
and 
does not saturate but increases to
a value over 1, indicating that the
vortices can diffuse more than a lattice constant 
in the $x$ direction over time.  
This is due to the formation
of dislocations which allow adjacent rows of vortices to slip past
each other while remaining confined in the $y$ direction.
We note that the saturation value of $d_{y}/a$ is approximately $1/5$, 
larger than the Lindemann criterion value of $1/10$. 
Excess motion in the $y$ direction occurs during a sliding event when two
rows slip past each other and the vortices in each row are 
temporarily displaced in the direction perpendicular to the slip plane.
This transverse motion is not large enough to permit the formation of
dislocations aligned in the hard direction.
In the nematic phase, shown in
Fig. 2(b) 
at $T = 1.35$,
$d_{x}$ still increases
more rapidly than $d_{y}$; however, the continuous increase of both quantities
indicates that the particles are diffusing throughout the
entire system.

We next consider the effect of random disorder by adding $N_p=2N_v$ randomly
located pinning sites to the same system studied in Fig.~1, and
then 
conducting a

\begin{figure}
\center{
\epsfxsize=3.5in
\epsfbox{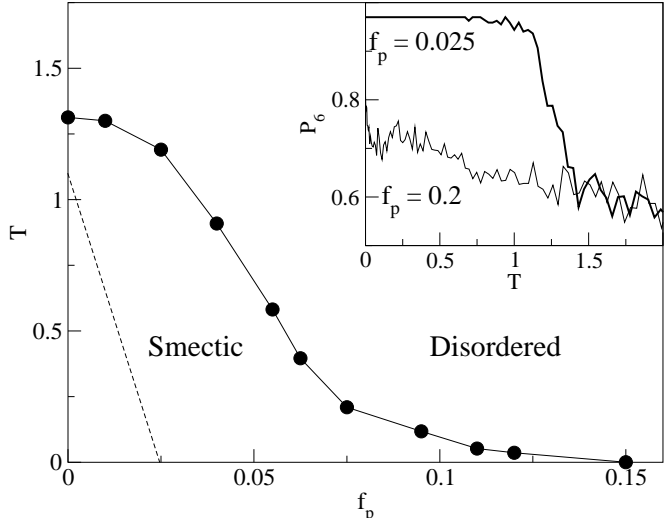}}
\caption{Regions in which the smectic and disordered phases
occur 
in a system with quenched disorder
as a function of temperature $T$ and pinning strength $f_{p}$. 
Dashed line roughly indicates the weak pinning region in which a 2D anisotropic
Bragg glass forms.
Inset: the density of six-fold coordinated particles 
$P_6$ vs $T$ for 
(top curve) $f_{p} = 0.025$ and (bottom curve) $f_{p} = 0.2$.    
}
\end{figure}

\noindent
series of simulations at varied $T$ and varied pinning strength $f_{p}$. 
For high temperatures we always
obtain a nematic phase, while at $T = 0$ for large $f_p$
we find a pinned nematic
phase. At low $T$ and low $f_{p}$ we observe 
a phase very similar to that shown in 
Fig.~1(c,d), where the vortex lattice is oriented in the soft 
direction and there are a small number of 
aligned dislocations.
We term this a pinned smectic-A phase.
In Fig.~3 we indicate the regions in which the smectic and disordered
phases appear as a function of temperature and pinning strength.
The phase boundary is identified 
via the density of sixfold coordinated particles, $P_6$; the defect
density is given by 1-$P_6$.  
In the crystal phase, there are no defects and $P_6=1$. In the
smectic phase, $P_6=0.91$ to $0.95$, and in the nematic phase 
$P_6>0.8$.
In the inset of Fig.~3 we plot 
$P_{6}$ vs $T$ 
for two different disorder strengths. 
For $f_{p} = 0.025$ (upper line) the
system is in the pinned smectic state at $T = 0$. 
As $T$ increases, there is a 
clear transition to the disordered state, as indicated
by the drop in in $P_{6}$ near $T = 1.19$.
The lower line shows $P_{6}$ 
for $f_p=0.2$, when the pinning is strong enough to disorder 
the system even at $T = 0$.  
These results suggest that weak random disorder can increase the
extent of the regions where the smectic-A phase occurs
when there are anisotropic interactions, by suppressing the crystalline
phase at low temperatures and raising the melting temperature of the
smectic state.    

It has been shown that dislocations are always present
for weak disorder in two dimensional isotropic systems; 
however, the distance between the dislocations can be arbitrarily large 
compared to the range of the 

\begin{figure}
\center{
\epsfxsize=3.5in
\epsfbox{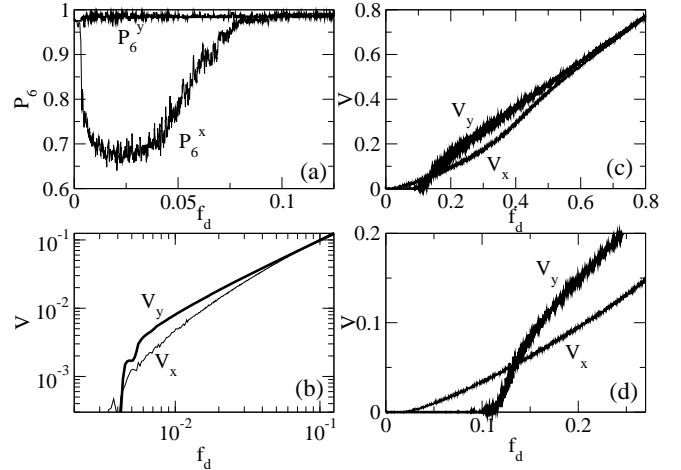}}
\caption{
(a) 
$P_{6}$ vs driving force $f_{d}$ for a
system with $f_{p} = 0.04$ at $T=0$. 
Upper curve: 
$P^{y}_{6}$ for ${\bf f}_{d}=f_d{\bf \hat{y}}$. 
Lower curve: 
$P^{x}_{6}$ for ${\bf f}_{d}=f_d{\bf \hat{x}}$.  
(b) Average velocities vs $f_{d}$ for the same system. 
Upper curve: 
$V_{y}$ for ${\bf f}_{d}=f_d{\bf \hat{y}}$. 
Lower curve:
$V_{x}$ for ${\bf f}_{d}=f_d{\bf \hat{x}}$.  
(c) $V_x$ and $V_y$ vs $f_{d}$ 
for a system with $f_{p} = 0.2$. 
(d) A blowup of (c) in the region near depinning 
showing the crossing of the velocity force 
curves.
}
\end{figure}

\noindent
translational order, 
so that for a wide range of temperatures and disorder strengths the
system behaves as a 2D Bragg glass \cite{Giamarchi}. 
In the vortex liquid crystal case there are
two length scales associated with the hard and soft directions, and 
the disorder induced dislocations first form in the soft direction.
It may be possible that, on very large
length scales, dislocations in the hard direction will also appear. 
For the parameters considered here,
dislocations are present except at the lowest pinning strengths, roughly
indicated by a dashed line in Fig. 3, 
where a 2D anisotropic Bragg glass forms. 

We next consider dynamical effects in the presence of pinning.
In the smectic-A state, 
there should be distinct
transport signatures for the hard and soft directions.
When the
disorder is strong enough to destroy the smectic phase, there may still
be an anisotropic transport signature
if the system retains some form of nematic order.  
We first consider the pinned smectic-A state found at
$f_{p} = 0.04$ and $T = 0$. 
We 
perform separate simulations for driving in the
soft direction, ${\bf f}_d=f_d{\bf \hat{x}}$, and the 
hard direction, ${\bf f}_d=f_d{\bf \hat{y}}$, 
increasing the applied drive very slowly to avoid any
transient effects. In Fig.~4(b) we plot 
$V_{y}=(1/N_v)<\sum_i^{N_v}v_y>$ (upper curve) for driving
in the hard direction and
$V_{x}=(1/N_v)<\sum_i^{N_v}v_x>$ (lower curve)
for driving in the soft direction. 
Here, $V_x<V_y$, 
indicating that motion in the soft direction is easier
except at
very high drives when the effects of the pinning are washed out and the two
curves come together.
In Fig.~4(a) we plot 
$P_{6}$ for the two different driving directions. 
At $f_{d} = 0,$ $P_{6}$ is slightly less than one due to the presence of 
a small number of dislocations in the smectic state. 
For depinning in the hard direction, $P_6^y$ (upper curve),
the system depins {\it elastically} without a proliferation of defects, and 
the vortices do not exchange neighbors as they move. 
For driving in the
easy direction, $P_{6}^x$ (lower curve) drops substantially when the 
vortices depin {\it plastically}, and
a portion of the vortices remain pinned while others flow past. 
The effective
pinning is known to be higher for a soft system where defects can proliferate
than for an elastic system
\cite{Higgins}. 
A similar proliferation of defects has been associated with
the so called peak effect, where there is a sudden increase in the 
effective pinning force as a function of temperature or applied
magnetic field
\cite{Higgins}.   
At high drive, the system shifts from plastic flow to a dynamically reordered
state 
\cite{Zimanyi} as indicated by the
increase in $P_{6}^x$ in Fig.~4(a), 
as well as by the merging of $V_{x}$ and $V_{y}$ in Fig.~4(b).
These results imply that, in the pinned smectic-A state, 
the depinning is 
elastic in the hard direction and plastic in the soft direction. 
Further, the elastic and plastic depinning transitions produce
different scaling responses in the
velocity force curves. At depinning, the velocity scales with the 
driving force in the
form $V = (f_{d} -f_{c})^{\beta}$ \cite{Fisher}. 
For the plastic flow regime we find $\beta > 1.0$ while
in the elastic flow regime we find $\beta < 1.0$, 
in agreement with theoretical expectations.

At finite temperatures and for $f_{p}$ large enough 
that we observe only plastic
depinning in both directions, we observe that the critical depinning force 
in the soft direction, $f_c^x$, is
{\it lower} than the depinning force in the hard direction, $f_c^y$,
even though $V_x<V_y$
at intermediate drives. 
This implies that the anisotropic flow
exhibits a reversal from $V_x>V_y$ to $V_x<V_y$ at low drives. 
We explicitly demonstrate this effect for a system with $f_p=0.25$ and
$T=0.25$ in Fig.~4(c,d).
Here, the
depinning is plastic in both directions, and the depinning forces are
$f_c^x=0.015$ and $f_c^y=0.11$. 
There are fewer dislocations for ${\bf f}_{d}=f_d{\bf \hat{y}}$
and the system reorders at $f_{d}^y = 0.2$. 
For 
${\bf f}_{d}=f_d{\bf \hat{x}}$, the
system does not reorder until $f_{d}^x = 0.8$. 
There is a clear crossing of the 
velocity force curves
at $f_{d} = 0.13$ so that the flow is easier in the soft direction 
for $f_d<0.13$ and easier in the hard direction for $f_d>0.13$.
In Fig. 4(d) we show a blowup of this region. 
The crossing of the velocity force curves can be understood 
by considering that the depinning in the
soft direction is plastic. At low drives, 
individual vortices can be thermally activated,
giving rise to creep. For driving in the hard direction, 
the depinning is elastic and individual vortex hopping is not possible,
so that only collective creep can occur.
In the case of the nematic phase, when there is some plastic flow in the 
$y$-direction, there is still
a large correlated length scale that must move so 
thermal effects are greatly reduced.  
Thus, creep in the pinned smectic phase and pinned nematic phase is 
enhanced in the soft direction
compared to the hard direction. 
Even for very high values of $f_{p}$ we observe an intermediate
anisotropic response, suggesting the system can be considered a 
pinned nematic phase. 

In conclusion, we have performed simulations of the recently proposed 
vortex liquid crystal state
where the vortex-vortex interactions are anisotropic. We find that, in 
the absence of disorder,
the system shows an intermediate melting into a smectic-A state 
as proposed theoretically in Ref.~\cite{Carlson}.
The smectic-A state contains 
a small fraction of dislocations which are all aligned in the soft
direction. 
In the presence of disorder, a pinned smectic-A state can occur, and
the system depins plastically in the soft direction
but elastically in the hard direction. 
We predict that, for equal intermediate
drives, the velocity in the soft direction will be lower than in 
the hard direction. At finite 
temperatures the creep is much higher in the soft direction due to the fact
that individual vortex hopping can occur,
whereas creep is suppressed in the hard direction since the vortex 
motion is much more correlated. 
For high temperatures and disorder strengths, the system
is disordered. 
For strong disorder the
anisotropic transport should still be observable in the pinned nematic 
phase. We note that
many of our results may apply to electron liquid crystals as well.  

We thank E. Carlson for useful discussions. 
This work was supported by the U.S.~Department of Energy under
Contract No.~W-7405-ENG-36.

%\vspace{-0.2in}

\end{document}